\documentclass[sn-mathphys,twocolumn,sn-basic,Numbered]{sn-jnl}

\usepackage{graphicx}%
\usepackage{multirow}%
\usepackage{amsmath,amssymb,amsfonts}%
\usepackage{amsthm}%
\usepackage{mathrsfs}%
\usepackage[title]{appendix}%
\usepackage{xcolor}%
\usepackage{textcomp}%
\usepackage{manyfoot}%
\usepackage{booktabs}%
\usepackage{algorithm}%
\usepackage{algorithmicx}%
\usepackage{algpseudocode}%
\usepackage{listings}%
\usepackage[permil]{overpic}
\usepackage{gensymb}
\usepackage{lineno}
\usepackage{lmodern}
\usepackage{anyfontsize}
\usepackage{verbatim}
\usepackage{amsmath}
\usepackage{cancel}

\usepackage{graphicx}
\usepackage{dcolumn}
\usepackage{bm}
\usepackage{xspace}
\usepackage{units}
\usepackage{setspace}

\usepackage{etoolbox}

\newcommand{\nue}{\ensuremath{\nu_{e}}\xspace}
\newcommand{\nueCC}{$\nu_{e}$\,CC}
\newcommand{\numu}{\ensuremath{\nu_{\mu}}\xspace}
\newcommand{\numuCC}{$\nu_{\mu}$\,CC}
\newcommand{\nutau}{\ensuremath{\nu_{\tau}}\xspace}
\newcommand{\pio}{\ensuremath{\pi^0}\xspace}

\setcitestyle{square,numbers}
\begin{document}
\makeatletter
\renewcommand\footnoterule{%
	\kern -3pt
	\hrule \@width \columnwidth
	\kern 2.6pt
}
\makeatother

\title{Search for Light Sterile Neutrinos With Two Neutrino Beams at MicroBooNE}

\author{The MicroBooNE Collaboration\footnotemark[2]}

\abstract{
The existence of three distinct neutrino flavours, \nue, \numu, and \nutau, is a central tenet of the Standard Model of particle physics~\cite{Glashow:1961tr,Weinberg:1967tq}.
Quantum-mechanical interference can allow a neutrino of one initial flavour to be detected some time later as a different flavour, a process called neutrino oscillation.
Several anomalous observations inconsistent with this three-flavour picture have motivated the hypothesis that an additional neutrino state exists which does not interact directly with matter, termed a ``sterile" neutrino, $\nu_s$~\cite{Aguilar:2001ty,Aguilar-Arevalo:2013pmq,Aguilar-Arevalo:2020nvw,Kaether:2010ag,SAGE:2009eeu,Barinov:2022wfh,Serebrov:2020kmd}. This includes anomalous observations from the LSND~\cite{Aguilar:2001ty} and MiniBooNE~\cite{Aguilar-Arevalo:2013pmq,Aguilar-Arevalo:2020nvw} experiments, consistent with $\numu\rightarrow\nue$ transitions at a distance inconsistent with the three-neutrino picture.
Here, we use data obtained from the MicroBooNE liquid-argon time projection chamber~\cite{Acciarri:2016smi} in two accelerator neutrino beams to exclude the single light sterile neutrino interpretation of the LSND and MiniBooNE anomalies at the 95\% confidence level (CL). Additionally, we rule out a significant portion of the parameter space that could explain the gallium anomaly~\cite{Kaether:2010ag,SAGE:2009eeu,Barinov:2022wfh}.
This is the first measurement to use two accelerator neutrino beams to break a degeneracy between \nue appearance and disappearance that would otherwise weaken the sensitivity to the sterile neutrino hypothesis.
We find no evidence for either $\numu\rightarrow\nue$ flavour transitions or \nue disappearance that would indicate non-standard flavour oscillations.
Our results show that previous anomalous observations consistent with $\numu\rightarrow\nue$ transitions cannot be explained by introducing a single sterile neutrino state. 
}

\maketitle

\begingroup
\renewcommand\thefootnote{\fnsymbol{footnote}}
\footnotetext[2]{A list of authors and their affiliations appears at the end of the paper.}
\endgroup

A broad experimental programme has shown that the three quantum mechanical eigenstates of neutrino flavour, $\nu_e$, $\nu_\mu$, and $\nu_\tau$, are related to the three eigenstates of neutrino mass, $\nu_{1}$, $\nu_{2}$, and $\nu_{3}$, by the unitary Pontecorvo-Maki-Nakagawa-Sakata (PMNS) matrix~\cite{Pontecorvo:1967fh, Maki:1962mu}.
This mixing between flavour and mass states gives rise to the phenomenon of neutrino oscillation, in which neutrinos transition between flavour eigenstates with a characteristic wavelength in $L/E_\nu\propto (\Delta m^2_{ji})^{-1}$, where $L$ is the distance travelled by the neutrino, $E_\nu$ is the neutrino energy, and $\Delta m^2_{ji}=m_j^2-m_i^2$ is the difference between the squared masses of the mass eigenstates $\nu_i$ and $\nu_j$.
The three known neutrino mass states give rise to two independent mass-squared differences, and thus to two characteristic oscillation frequencies that have been well measured with neutrinos from nuclear reactors~\cite{ref:KamLANDThreeFlavour,ref:DayaBayThreeFlavour}, the Sun~\cite{ref:SNOThreeFlavour}, the Earth's atmosphere~\cite{Super-Kamiokande:1998kpq,ref:DeepCoreThreeFlavour}, and particle accelerators~\cite{ref:MINOSThreeFlavour,ref:NOvAThreeFlavour,ref:T2KThreeFlavour}.

In apparent conflict with the three-neutrino model, several experiments during the past three decades have made observations that can be interpreted as neutrino flavour-change with a wavelength much shorter than is possible given only the two measured mass-squared differences~\cite{Aguilar:2001ty,Aguilar-Arevalo:2013pmq,Aguilar-Arevalo:2020nvw,Kaether:2010ag,SAGE:2009eeu,Barinov:2022wfh,Serebrov:2020kmd}.
These observations are often explained as neutrino oscillations caused by at least one additional mass state, $\nu_4$, corresponding to a mass-squared splitting of $\Delta m^2_{41} \gtrsim 10^{-2}$eV$^2$, which is much greater than the measured $\Delta m^2_{21}$ and $\Delta m^2_{32}$.
New mass states would require the addition of an equivalent number of new flavour states, in conflict with measurements of the $Z$-boson decay width~\cite{ALEPH:2005ab} that have definitively shown that only three light neutrino flavour states couple to the $Z$ boson of the weak interaction. 
Therefore, these additional neutrino flavour states must be unable to interact through the weak interaction and are thus referred to as ``sterile" neutrinos. In this analysis we focus specifically on light sterile neutrinos - those with masses below at least half the mass of the $Z$ boson. It should be noted that the term ``sterile neutrino" has also been used to describe new particles, such as heavy right-handed lepton partners, that are potentially more massive than the $Z$ boson; however, our study does not directly test such scenarios.
The discovery of additional neutrino states would have profound implications across particle physics and cosmology, for example on our understanding of the origin of neutrino mass, the nature of dark matter, and the number of relativistic degrees of freedom in the early universe.

With the addition of a single new mass state $\nu_4$ and a single sterile flavour state $\nu_s$, the PMNS matrix becomes a $4\times 4$ unitary matrix described by six real mixing angles $\theta_{ij}$ ($1\leq i<j\leq 4$).
Oscillations driven by the two measured mass-squared splittings have not had time to evolve for small values of $L/E_{\nu}$. The \numu to \nue flavour-change probability, $P_{\nu_\mu\rightarrow\nu_e}$, and the \nue and \numu survival probabilities, $P_{\nu_e\rightarrow\nu_e}$ and $P_{\nu_\mu\rightarrow\nu_\mu}$, can then, to a very good approximation, be described by
\begin{eqnarray}
    P_{\nu_\mu\rightarrow\nu_e} &=& \sin^2(2\theta_{\mu e})\sin^2\left(\frac{\Delta m^2_{41}L}{4E_\nu}\right),\label{eq:app_oscprob}\\
    P_{\nu_e\rightarrow\nu_e}
    &=& 1 - \sin^2(2\theta_{ee})\sin^2\left(\frac{\Delta m^2_{41}L}{4E_\nu}\right),\label{eq:disapp_oscprobE}\\
    P_{\nu_\mu\rightarrow\nu_\mu}
    &=& 1 - \sin^2(2\theta_{\mu\mu})\sin^2\left(\frac{\Delta m^2_{41}L}{4E_\nu}\right),\label{eq:disapp_oscprobMu}
\end{eqnarray}
where $\theta_{ee}\equiv\theta_{14}$, $\sin^2(2\theta_{\mu\mu})\equiv4\cos^2\theta_{14}\sin^2\theta_{24}(1-\cos^2\theta_{14}\sin^2\theta_{24})$, and $\sin^2(2\theta_{\mu e})\equiv\sin^2(2\theta_{14})\sin^2\theta_{24}$, following the common parameterization~\cite{Giunti:2019aiy}. 
Flavour transitions due to these new oscillation parameters are experimentally probed by observing unexpected deficits or excesses in charged current (CC) \nue\ and \numu interactions in a flavour-sensitive neutrino detector from a source of well-defined neutrino flavour content.  

Observations compatible with a fourth neutrino mass state have been made in measurements of intense electron-capture decay sources~\cite{Kaether:2010ag,SAGE:2009eeu,Barinov:2022wfh}, where a deficit in detected $\nu_e$ rates implies non-unity $P_{\nu_e\rightarrow\nu_e}$ from a $\Delta m^2_{41}>\mathcal{O}(\unit[1]{eV^2})$.   
While a hint of non-unity $P_{\,\overline{\nu}_e\rightarrow\overline{\nu}_e}$ is provided by the nuclear reactor-based Neutrino-4 experiment~\cite{Serebrov:2020kmd}, this result is in conflict with other reactor-based observations from DANSS, NEOS, PROSPECT, and STEREO, which see no evidence for $L/E_\nu$-dependent $\overline{\nu}_e$ disappearance~\cite{Danilov:2021oop,RENO:2020hva,Andriamirado:2024glk,STEREO:2019ztb}.  
Two accelerator-based experiments, LSND and MiniBooNE, have observed potential evidence of non-zero $P_{\nu_\mu\rightarrow\nu_e}$ associated with large mass splittings of $\Delta m^2_{41}>\mathcal{O}(\unit[10^{-2}]{eV^2})$.  
The LSND experiment observed an anomalous excess of $\overline{\nu}_e$ interactions in a $\pi^+$ decay-at-rest beam~\cite{Aguilar:2001ty}.
The MiniBooNE experiment, situated downstream from the Booster Neutrino Beam (BNB) proton target facility generating a beam of GeV-scale $\nu_{\mu}$ and $\overline{\nu}_\mu$ from decays of boosted $\pi^+$ and $\pi^-$, observed an excess of electromagnetic showers indicative of $\nu_e$ interactions that would imply a non-zero $P_{\nu_\mu\rightarrow\nu_e}$~\cite{Aguilar-Arevalo:2013pmq,Aguilar-Arevalo:2020nvw}.  
Observations of \nue\ disappearance and \nue\ appearance should be accompanied by \numu\ disappearance (non-unity $P_{\nu_\mu\rightarrow\nu_\mu}$) if the PMNS matrix is unitary.
No conclusive observation of such \numu disappearance has been reported~\cite{ref:MINOSSterile,ref:IceCubeSterile, PhysRevLett.134.081804}.
The overall picture of the existence and phenomenology of sterile neutrino states thus remains inconclusive.

In this article, we present new results on sterile neutrino oscillations from the MicroBooNE liquid-argon time-projection chamber (LArTPC) experiment at Fermilab~\cite{Acciarri:2016smi}.  
Situated along the same BNB beamline hosting the MiniBooNE experiment, MicroBooNE was conceived to directly test MiniBooNE's non-zero $P_{\nu_\mu\rightarrow\nu_e}$ observation.
By supplanting MiniBooNE's Cherenkov detection technology with the precise imaging and calorimetric capabilities of a LArTPC, MicroBooNE can reduce backgrounds and select a high purity sample of true $\nu_e$-generated final-state electrons. 
MicroBooNE's first $\nu_e$ measurement results using differing final-state topologies showed no evidence for an excess of $\nu_e$-generated electrons from the BNB~\cite{MicroBooNE:2021ktl,MicroBooNE:2021bcu,MicroBooNE:2021pld,MicroBooNE:2021nxr}.  
These results were used to set limits on $\nu_\mu\rightarrow\nu_e$ flavour transitions, excluding sections of the region in $(\Delta m^2_{41},\sin^2(2\theta_{\mu e}))$ space favoured by LSND and MiniBooNE data~\cite{WCOscPRL}. 
Since the BNB has an intrinsic contamination of electron neutrinos, the disappearance of electron neutrinos can cancel the appearance of electron neutrinos from $\nu_\mu\rightarrow\nu_e$ oscillations~\cite{Arguelles:2021meu}. 
This effect leads to a degeneracy between the impact of the mixing angles $\theta_{\mu e}$ and $\theta_{ee}$ of Eqs.~\ref{eq:app_oscprob} and~\ref{eq:disapp_oscprobE} that weakens the sensitivity to the parameters of the expanded $4\times 4$ PMNS matrix.

\begin{figure}
	\centering
	\begin{overpic}[width=\columnwidth]{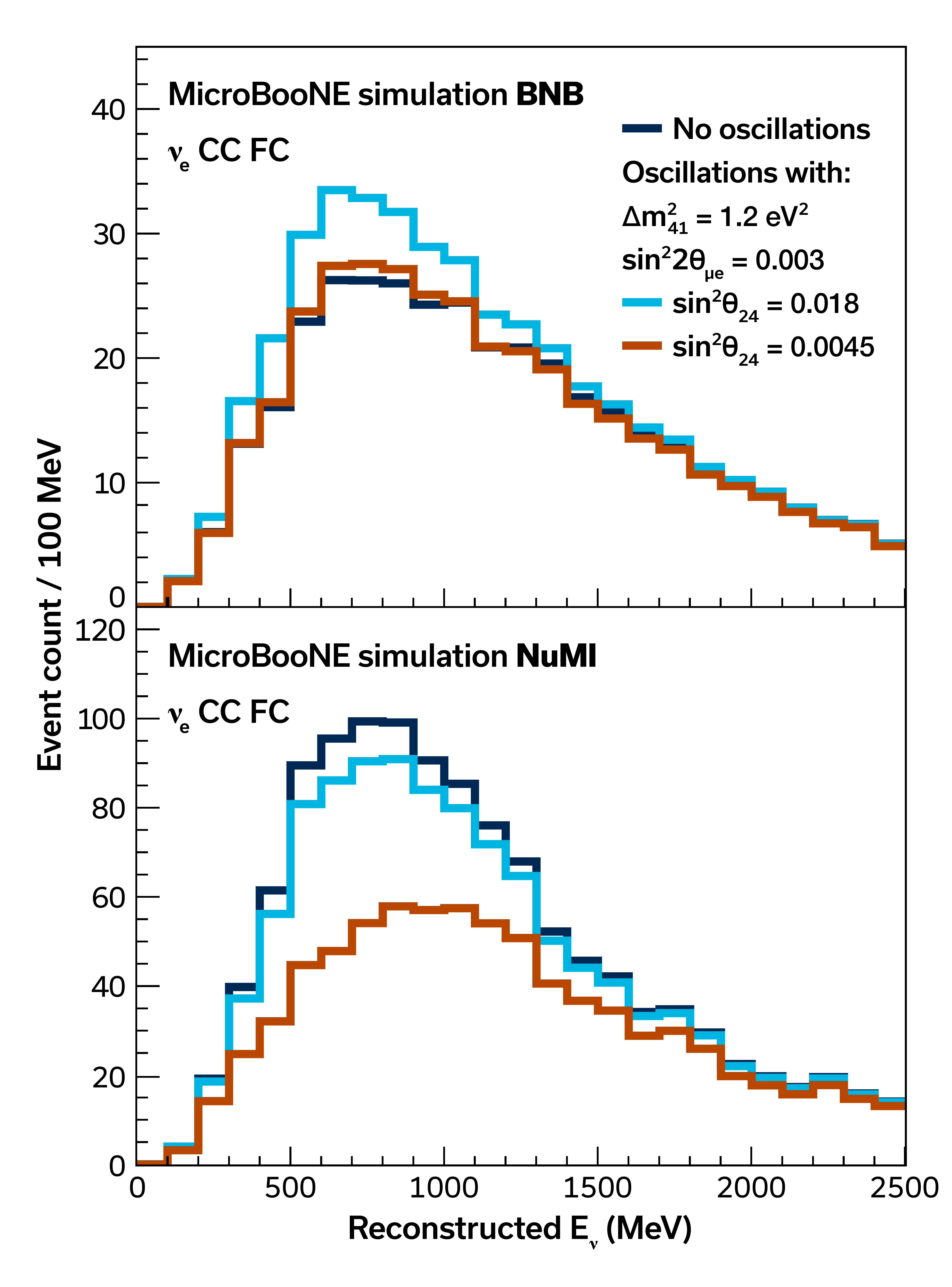}
		\put(675,935){\bf a}
		\put(675,495){\bf b}
	\end{overpic}
	\caption{{\bf Illustration of the breaking of the degeneracy between $\nu_e$ appearance and disappearance enabled by the independent BNB and NuMI data sets in MicroBooNE.}
		Simulated reconstructed energy spectra of fully contained (FC) charged current (CC) $\nu_e$ interactions in MicroBooNE from ({\bf a}) the BNB and ({\bf b}) the NuMI beam.
		The dark blue histograms show the $3\nu$ expectation for $\sin^2\theta_{24}=\sin^2(2\theta_{\mu e})=0$.
		The light blue and red histograms show expectations for two sets of parameters of the $4\nu$ model, both with $\Delta m^2_{41}=\unit[1.2]{eV^2}$ and $\sin^2(2\theta_{\mu e})=0.003$ (corresponding . 
		The light blue histograms show the expectation for $\sin^2\theta_{24}=0.018$ and the red histograms show the expectation for $\sin^2\theta_{24}=0.0045$. Note that these parameters were chosen specifically to highlight differences in the oscillated spectra between BNB and NuMI, and do not imply that parameter spaces associated with these values are newly excluded by this result.}
	\label{fig:DegeneracySpectra}
\end{figure}

We overcome the limitations of the degeneracy between $\nu_e$ appearance and $\nu_e$ disappearance by performing the first-ever oscillation search using two accelerator neutrino beams: the BNB and the Neutrinos at the Main Injector (NuMI) beam. 
The MicroBooNE detector is aligned with the BNB's direction and is at an angle of about 8$^{\circ}$ relative to the NuMI beam. Beam timing information is utilized to distinguish and record events from each beam separately.
This configuration results in two neutrino data sets differing in intrinsic electron flavour fraction. The BNB's electron flavour content is 0.57\% and that of the NuMI beam is 4.6\%. 
These two independent sets of data, with substantially different electron flavour contents, break the degeneracy between $\nu_e$ appearance and disappearance.
We illustrate the impact of using two beams in Fig.~\ref{fig:DegeneracySpectra}, where we compare simulated $\nu_e$ energy spectra from the BNB and the NuMI beam for the three-flavour ($3\nu$) hypothesis and for two sets of parameters of the expanded four-flavour ($4\nu$) PMNS model with $\Delta m^2_{41}=\unit[1.2]{eV^2}$ and $\sin^2(2\theta_{\mu e})=0.003$.
For $\sin^2\theta_{24}=0.0045$ the $\nu_e$ appearance and disappearance cancel in the BNB leaving a $\nu_e$ spectrum that is almost identical to the $3\nu$ case, whereas the NuMI beam shows a clear indication of $\nu_e$ disappearance.
The appearance and disappearance effects almost fully cancel in the NuMI beam for $\sin^2\theta_{24}=0.018$ while the BNB shows a clear indication of $\nu_e$ appearance. 
In the Methods section, we provide further discussion of this degeneracy over a broader range of mass-squared splittings and mixing angles.

\begin{table*}[ht]
	\centering
	\begin{tabular}{c|cc|cc}
		\hline
		\hline
		& BNB FC & BNB PC & NuMI FC & NuMI PC \\ \hline
		\multicolumn{5}{c}{\bf Predicted and observed events in CC $\nu_e$ signal samples} \\ \hline
		Unconstrained true CC $\nu_e$ & 310.9 & 175.7 & 1107.7 & 554.2 \\
		Constrained true CC $\nu_e$ signal & 346.9 & 188.2 & 1446.5 & 703.1 \\
		Background & 53 & 40.7 & 79.2 & 104.7 \\ \hline
		Constrained total CC $\nu_e$ samples & 399.9 & 228.9  & 1525.7 & 807.8 \\ \hline
		Data & 338 & 219 & 1490 & 824 \\ \hline
		\multicolumn{5}{c}{\bf Systematic uncertainties on CC $\nu_e$ signal samples} \\ \hline
		Neutrino flux prediction & 5.9\% & 6.1\% & 19.6\% & 19.7\% \\
		Neutrino interaction uncertainties & 14.7\% & 14.0\% & 17.5\% & 15.1\% \\
		Detector uncertainties & 3.3\% & 3.2\% & 2.0\% & 3.9\% \\
		Monte Carlo statistics & 1.57\% & 1.95\% & 1.16\% & 1.67\% \\ \hline
		Total unconstrained uncertainty & 16.3\% & 15.8\% & 26.4\% & 25.2\% \\ 
		Total constrained uncertainty & 4.5\% & 5.5\% & 5.8\% & 5.9\% \\ \hline\hline
		
	\end{tabular}
	\caption{{\bf Event counts and systematic uncertainties.}
		The predicted and observed events counts, and the unconstrained and constrained systematic uncertainties, on the fully contained (FC) and partially contained (PC) CC $\nu_e$ samples from the BNB and the NuMI beam. All the expected numbers correspond to 6.369$\times$10$^{20}$ POT for BNB and 10.54$\times$10$^{20}$ POT for NuMI.}
	\label{tab:uncertainties}
\end{table*}

Using the two-beam technique, this new MicroBooNE analysis achieves significant improvements in sensitivity to the parameters $\sin^2(2\theta_{ee})$ and $\sin^2(2\theta_{\mu e})$ relative to MicroBooNE's prior sterile neutrino analysis over a broad range of $\Delta m^2_{41}$ values. 
These improvements are shown by the sensitivities presented in Extended Data Fig.~\ref{extfig:Sensitivities}.
The results presented here using two neutrino beams place robust new constraints on the validity of the sterile neutrino hypothesis in explaining existing short-baseline anomalies in neutrino physics.  
This analysis strengthens MicroBooNE's direct test of the sterile-neutrino interpretation of the MiniBooNE anomaly and allows MicroBooNE to probe the $\sin^2(2\theta_{\mu e})$ parameter space favoured by LSND.
We also constrain $\sin^2(2\theta_{ee})$, complementing existing exclusions from reactor, solar~\cite{Goldhagen:2021kxe,Berryman:2021yan}, and $\beta$-decay~\cite{KATRIN:2022ith} experiments, thereby further restricting the sterile neutrino parameter space relevant to the gallium anomaly.

We use data corresponding to $6.369\times 10^{20}$\,protons on target (POT) in the BNB, with magnetic van-der-Meer horns configured to focus positively charged hadrons leading to a $\nu_\mu$-dominated beam with a 5.9\% $\overline{\nu}_\mu$ component and a 0.57\% $\nu_e+\overline{\nu}_e$ component.  
From the NuMI beam, a total of $\unit[10.54\times 10^{20}]{POT}$ are used, where $30.8\%$ were taken with horns configured to focus positively charged hadrons and the remainder with horns focusing negatively charged hadrons.
The NuMI flux observed in the MicroBooNE detector, with both horn configurations combined, is $\nu_\mu$ dominated with a 42.1\% $\overline{\nu}_\mu$ component and a 4.6\% $\nu_e+\overline{\nu}_e$ component.
In the rest of this article, we do not discriminate between neutrinos and antineutrinos and refer to the $\nu_\mu+\overline{\nu}_\mu$ and $\nu_e+\overline{\nu}_e$ samples as $\nu_\mu$ and $\nu_e$ samples for brevity. For both BNB and NuMI, the POT used in this analysis represent roughly half of the total data collected by the MicroBooNE detector; additional data remain available for future studies.

MicroBooNE's LArTPC detector has an active volume of $\unit[10.4\times 2.6\times 2.3]{m^3}$ containing \unit[85]{tonnes} of liquid argon.  
Charged particles passing through the argon create ionisation trails.
A $\unit[273]{V/cm}$ electric field drifts the ionisation electrons towards an anode plane consisting of three layers of wires separated by \unit[3]{mm} and each with a \unit[3]{mm} wire pitch that collects the electrons and enables three-dimensional imaging of the neutrino interactions. 
The passage of charged particles through the argon also produces scintillation light that is collected by a system of photomultiplier tubes to provide timing information.  
Signal processing and calibrations of MicroBooNE data are described in Ref.~\cite{ref:MicroBooNE:NoiseFiltering,ref:MicroBooNE:SignalProcessing1,ref:MicroBooNE:SignalProcessing2,ref:MicroBooNE:ChargeLossCalibration,ref:MicroBooNE:Laser,ref:MicroBooNE:SpaceCharge}.
 
Neutrino interactions in the LArTPC are reconstructed with the Wire-Cell analysis framework~\cite{MicroBooNE:2021ojx}.
The techniques for identifying and reconstructing neutrino interactions and their energies have been described elsewhere~\cite{MicroBooNE:2021nxr}.
We select a sample of CC $\nu_e$ interactions from the BNB (NuMI beam) with 82\% (91\%) purity and 46\% (42\%) efficiency, and a sample of CC $\nu_\mu$ interactions with 92\% (78\%) purity and 68\% (62\%) efficiency.
The CC $\nu_e$ and CC $\nu_\mu$ samples are divided into fully contained (FC) and partially contained (PC) samples, depending on whether all charge depositions are contained in a fiducial volume \unit[3]{cm} within the TPC boundary. 
The CC \numu\ events that contain a reconstructed \pio\ are separated into two additional FC and PC samples per beam.
Neutral current (NC) interactions that produce a $\pi^0$ are distinguished by the absence of a long muon-like track and the presence of detached reconstructed electromagnetic showers.
These form an additional sample.
In total, we define 14 distinct event categories, seven for each beam.

We produce a Monte Carlo prediction of our 14 samples, to which we compare the data.
There is significant systematic uncertainty creating this Monte Carlo simulation.
The uncertainty on the predicted rates of the 14 samples is given in Tab.~\ref{tab:uncertainties} and is referred to as the unconstrained systematic uncertainty.
The largest uncertainties come from neutrino interaction modelling for the BNB samples, and from a combination of neutrino flux and interaction uncertainties for the NuMI samples. 
Many of these uncertainties are highly correlated.
Thus, a combined fit of all samples effectively constrains the uncertainties on the CC $\nu_e$ prediction and at the same time allows the CC $\nu_e$ prediction to be modified, as can be seen from Tab.~\ref{tab:uncertainties}.
The pionless samples constrain uncertainties on CC \nue signal events, while the \pio samples constrain uncertainties on the dominant background.

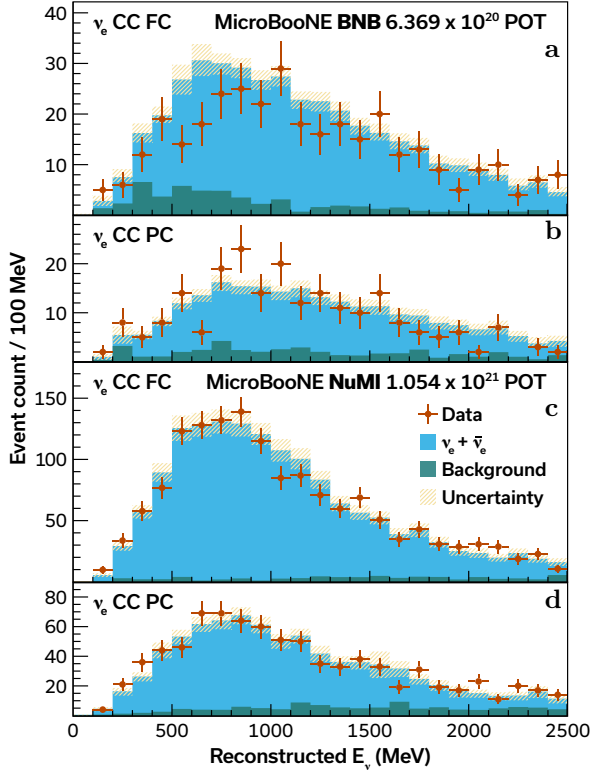
\begin{figure}
	\centering
	\begin{overpic}[width=\columnwidth]{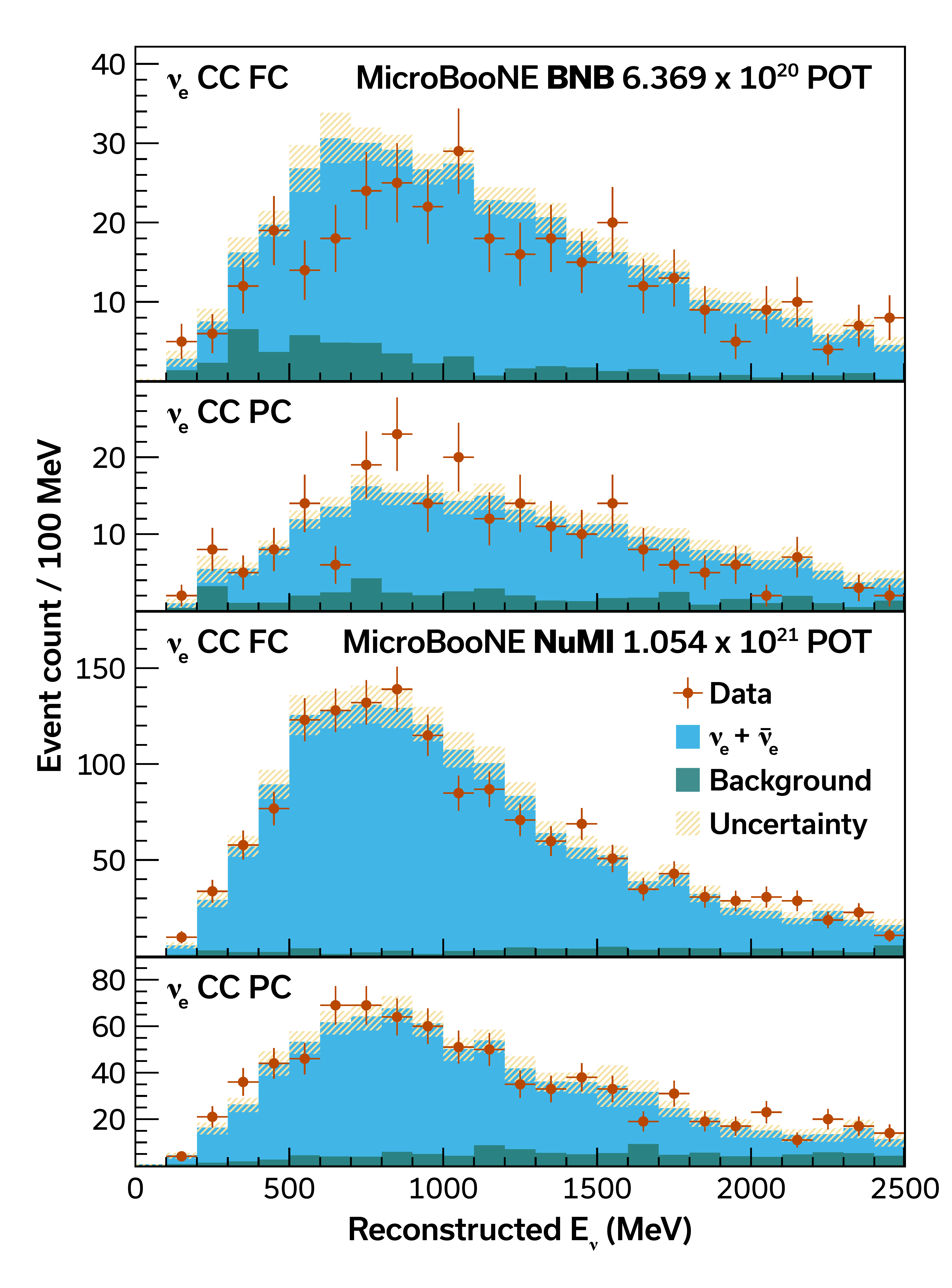}
		\put(675,897){\bf a}
		\put(675,675){\bf b}
		\put(675,460){\bf c}
		\put(675,226){\bf d}
	\end{overpic}
	\caption{{\bf Observed CC $\nu_e$ candidate events.} 
		Reconstructed energy spectra of events selected as ({\bf a}) fully contained CC $\nu_e$ candidates in the BNB, ({\bf b}) partially contained CC $\nu_e$ candidates in the BNB, ({\bf c}) fully contained $\nu_e$ candidates in the NuMI beam, and ({\bf d}) partially contained $\nu_e$ candidates in the NuMI beam.
		The data points are shown with statistical error bars.
		The constrained predictions for each sample are shown for the $3\nu$ hypothesis as the solid histograms, with the blue showing the true CC $\nu_e$ events and the green showing the background events.
		The background category contains CC $\nu_\mu$ interactions, NC neutrino interactions, cosmic rays, and interactions occurring outside the fiducial volume of the detector.
		The yellow band shows the total constrained systematic uncertainty on the prediction.}
	\label{fig:data}
\end{figure}

Uncertainties on the neutrino flux prediction arise from uncertainties in the production of charged pions and kaons in the BNB and NuMI targets and the material around the target halls and hadron-decay volumes.
These uncertainties are evaluated through comparison with external hadron production data~\cite{na49k,na49pi,na61}, following a procedure similar to that described in Ref.~\cite{ppfx}.  
The $\nu_e$ flux from three-body $K$ and $\mu$ decays is highly correlated with the $\nu_\mu$ flux from two-body $\pi$ and $K$ decays, allowing our $\nu_\mu$ samples to effectively constrain the uncertainties on the $\nu_e$ flux predictions.
The neutrino interaction model is tuned using datasets of pionless CC interactions from the T2K experiment~\cite{MicroBooNE:2021ccs}. Uncertainties on this neutrino interaction model are evaluated by varying the input parameters within their allowed uncertainties.
These uncertainties are correlated between the BNB and NuMI datasets, and between the CC $\nu_\mu$ and $\nu_e$ samples due to lepton universality of the weak interaction. 
Uncertainties on the simulation of the detector include uncertainties on the response of the detector to ionisation, uncertainties on the amount of ionisation charge freed by passing charged particles through the detector, uncertainties on the electric field map of the TPC, uncertainties on the production and propagation of scintillation light, uncertainties on backgrounds from interactions occurring outside the cryostat, and uncertainties on finite statistics of the simulation samples used for predictions.

The simultaneous fit to the 14 samples from the BNB and the NuMI beam incorporates all sources of systematic uncertainty through a covariance matrix.
We allow $\sin^2(2\theta_{\mu e})$, $\sin^2(2\theta_{ee})$, and $\Delta m^2_{41}$ complete freedom within unitarity bounds as parameters of the fit.
The covariance-matrix formalism $\chi^2$ test of the fit can be found in the Methods Section.
The constrained predictions shown in Fig.~\ref{fig:data} assume the $3\nu$ hypothesis of $\sin^2(2\theta_{\mu e})=\sin^2(2\theta_{ee})=0$.
They agree well with the data with a $p$-value of 0.92. The best-fit values for the oscillation parameters in the $4\nu$ hypothesis are $\Delta m^2_{41}=\unit[1.30\times 10^{-2}]{eV^2}$, $\sin^2(2\theta_{\mu e}) = 0.999$, and $\sin^2(2\theta_{ee})=0.999$, with a $\chi^2$ difference with respect to the $3\nu$ hypothesis of
\begin{linenomath*}
\begin{equation}
\Delta\chi^2=\chi^2_{\mathrm{null},3\nu}-\chi^2_{\mathrm{min},4\nu}=0.228.
\end{equation}
\end{linenomath*}  
We observe no significant preference for the existence of a sterile neutrino with a $p$-value of 0.96 evaluated using the Feldman-Cousins procedure.

\begin{figure}
	\centering
	\begin{overpic}[width=\columnwidth]{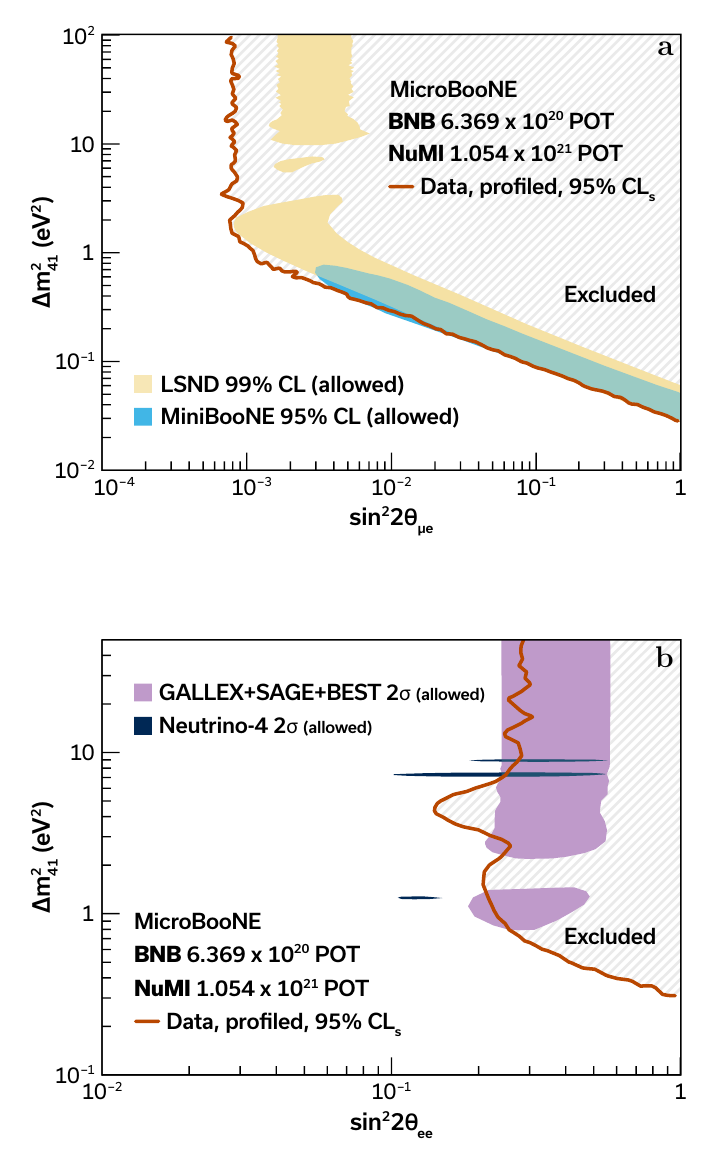}
	\end{overpic} 
	\caption{{\bf Constraints on parameters of the $4\nu$ oscillation model.}
		The red lines show exclusion limits at the 95\% $\mathrm{CL_{s}}$ level in the plane of $\Delta m^{2}_{41}$ and (\textbf{a}) $\sin^{2}(2\theta_{\mu e})$ or (\textbf{b}) $\sin^{2}(2\theta_{ee})$. 
		All the regions to the right of these lines are excluded by the MicroBooNE data.
		In (\textbf{a}), the yellow shaded area is the LSND  $99$\% CL allowed regions~\cite{Aguilar:2001ty} which neglects the degeneracy between $\nu_e$ disappearance and appearance. 
		The light blue area is the MiniBooNE 95\% CL allowed region~\cite{miniboone3p1} considering both $\nu_e$ disappearance and appearance.
		In (\textbf{b}), the purple shaded area is the 2$\sigma$ allowed region of the gallium anomaly~\cite{Barinov:2021asz}.
		The dark blue shaded area is the 2$\sigma$ allowed region from the \mbox{Neutrino-4} experiment~\cite{Serebrov:2020kmd}.
		For context, note that the stronger than expected constraint on $\sin^{2}(2\theta_{\mu e})$, driven by the deficit observed in the BNB \nueCC\ FC sample and the excess in the NuMI \numuCC\ sample, is discussed in detail in the Methods section and Extended Data Fig.~\ref{extfig:Sensitivities}.}
	\label{fig:Contours}
\end{figure}

Exclusion contours are calculated using the frequentist CL$_\mathrm{s}$ method~\cite{Read:2002hq}.
The exclusion contour in any two-dimensional parameter space is obtained by profiling the third free parameter.
At any point in the two-dimensional space, the value of the profiled parameter that minimises the $\chi^2$ with respect to the data is chosen. 
Fig.~\ref{fig:Contours}a shows the 95\% CL$_\mathrm{s}$ exclusion contour in the $(\Delta m^2_{41},\sin^2(2\theta_{\mu e}))$ parameter space.
The region allowed at 99\% CL by the LSND measurement and the vast majority of the region allowed at the 95\% CL by the MiniBooNE experiment are excluded.
Fig.~\ref{fig:Contours}b shows the 95\% CL$_\mathrm{s}$ exclusion contour in the $(\Delta m^2_{41},\sin^2(2\theta_{ee}))$ parameter space.
A significant portion of the region allowed by gallium measurements and part of the region derived from the Neutrino-4 measurement are excluded.
In the Methods Section and Extended Data Fig.~\ref{extfig:Sensitivities}, we compare our exclusions with the expected median sensitivities.

In summary, using data from the MicroBooNE detector, we report the first ever search for a sterile neutrino using two accelerator neutrino beams.
The oscillation fit to the $4\nu$ model using a total of 14 CC $\nu_e$, CC $\nu_\mu$, and NC $\pi^0$ samples from the BNB and the NuMI beam in a single detector achieves a significant reduction of systematic uncertainties and a powerful mitigation of degeneracies between $\nu_e$ appearance and disappearance.
The result shows no evidence of oscillations induced by a single sterile neutrino and is consistent with the $3\nu$ hypothesis with a $p$-value of 0.96.
We comprehensively exclude at 95\% confidence level the $4\nu$ parameter space that would explain the LSND and MiniBooNE anomalies through the existence of a light sterile neutrino in a model with an extended $4\times4$ PMNS matrix.
Our result expands the diverse range of experimental approaches excluding regions that would explain the gallium anomaly and the Neutrino-4 observation with a light sterile neutrino.
This work therefore provides a robust exclusion of a single light sterile neutrino as an explanation for the array of short-baseline neutrino anomalies observed over the past three decades, representing the strongest constraint from a short-baseline experiment using accelerator-produced neutrinos.
Expanded models including several light sterile neutrinos~\cite{Fong:2017gke}, neutrino decay effects~\cite{deGouvea:2019qre,Hostert:2024etd}, or production and decay of new particles connected with the dark  sector~\cite{Chang:2021myh,Dutta:2021cip} might explain the anomalies.
The Short Baseline Neutrino (SBN) Programme~\cite{ref:SBNProgramme} at Fermilab adds two new LArTPC detectors in the BNB, at different distances from the proton target. 
Future measurements by MicroBooNE and the broader SBN Programme can shed light on this expanded model space, with future comprehensive insights provided by near-term short-baseline measurements from diverse flavour channels and energy regimes.


\vspace{0.5cm}

\textbf{Acknowledgments} This document was prepared by the MicroBooNE collaboration using the resources of the Fermi National Accelerator Laboratory (Fermilab), a U.S. Department of Energy, Office of Science, HEP User Facility. Fermilab is managed by Fermi Research Alliance, LLC (FRA), acting under Contract No. DE-AC02-07CH11359.  MicroBooNE is supported by the following: the U.S. Department of Energy, Office of Science, Offices of High Energy Physics and Nuclear Physics; the U.S. National Science Foundation; the Swiss National Science Foundation; the Science and Technology Facilities Council (STFC), part of the United Kingdom Research and Innovation; the Royal Society (United Kingdom); the UK Research and Innovation (UKRI) Future Leaders Fellowship; and the NSF AI Institute for Artificial Intelligence and Fundamental Interactions. Additional support for the laser calibration system and cosmic ray tagger was provided by the Albert Einstein Center for Fundamental Physics, Bern, Switzerland. We also acknowledge the contributions of technical and scientific staff to the design, construction, and operation of the MicroBooNE detector as well as the contributions of past collaborators to the development of MicroBooNE analyses, without whom this work would not have been possible. All Transfer of Copyright Agreements must be signed by the Fermilab Technical Publications Office, including Creative Commons Licenses (CC-BY). This is required by our contract with DOE and gives us the right to make the information contained in the Scientific/Technical Publication freely available on Fermilab's Technical Publications web pages without breaking U.S. copyright laws. 

\vspace{0.5cm}

\textbf{Author contributions} The MicroBooNE Collaboration contributed collectively to this publication through the design, construction, and installation of the detector, its operation and data acquisition, and the development of simulation and analysis tools. The scientific results presented here were reviewed and approved by the entire collaboration.

\vspace{0.5cm}

\textbf{Competing interests} The authors declare no competing interests.

\vspace{0.5cm}

\textbf{Data availability} The measured data, predicted signal and background, along with their complete systematic uncertainties for the corresponding reconstructed neutrino energy bins in the \nue channels, are publicly accessible on \href{https://www.hepdata.net/record/166435?version=1&table=Unconstrained%2014%20channels}{\textsc{HEPData}} and \href{10.5281/zenodo.17161263}{\textsc{Zenodo}}. Additionally, $\Delta \chi^2$ values for each 4$\nu$ hypothesis across the three-dimensional grid of oscillation parameters are provided.

\vspace{0.5cm}

\textbf{Code availability} The MicroBooNE Collaboration is responsible for developing and maintaining the code used for simulating and analyzing the raw data that supports this result. This code is accessible to collaboration members but is not publicly available. Questions about the algorithms and methods employed in this analysis can be directed to the corresponding author.

\clearpage
\section*{Methods}
\setcounter{figure}{0}

\subsection*{Neutrino beams at MicroBooNE}
The BNB and the NuMI are conventional neutrino beamlines that use intense proton beam pulses to generate on-axis neutrino fluxes peaking in the neutrino-energy range 0.5--5 GeV.  
Boosted charged mesons escaping the target are focused into a decay pipe using magnetic horns, allowing decaying mesons to impart much of their kinetic energy directly to their neutrino product.  
The MicroBooNE detector is on the axis of the BNB, \unit[468.5]{m} from the proton target. The detector is about $8^\circ$ off-axis and \unit[679]{m} from the proton target of the NuMI beam.  
The muon flavour components of these beams are mostly generated via the primary decay channel of the dominant $\pi$ mesons, $\pi^\pm \rightarrow \mu^\pm + \nu_\mu(\overline{\nu}_\mu)$, while the electron flavour component is generated by decay of the pion's boosted $\mu^\pm$ daughter, via $\mu^\pm \rightarrow e^\pm + \overline{\nu}_\mu(\nu_\mu) + \nu_e(\overline{\nu}_e)$, and by semileptonic decays of sub-dominant $K$ mesons, specifically $K^\pm \rightarrow \pi^0 + e^\pm + \nu_e(\overline{\nu}_e)$ and $K^0_L \rightarrow \pi^\mp + e^\pm + \nu_e(\overline{\nu}_e)$.
  
Neutrinos in the BNB are created by colliding protons with a kinetic energy of \unit[8]{GeV} on a beryllium target, while in the NuMI beamline, \unit[120]{GeV} protons collide with a carbon target.
These differences serve to generate a higher on-axis beam energy in the NuMI beam, as well as a greater proportion of $K$ meson production, leading to a higher $\nu_e$ content at MicroBooNE's highly off-axis position.  
The NuMI beam also incorporates a longer charged-meson decay pipe (\unit[675]{m}) than the BNB (\unit[50]{m}), which increases its electron-flavour content by facilitating a higher proportion of decays of secondary $\mu^+$ from upstream $\pi^+$ decay.  
While the latter effect drives higher electron-flavour content on-axis for NuMI relative to the BNB, it is NuMI's larger proportion of unfocused or poorly focused $K$ mesons that drives its elevated electron-flavour content at MicroBooNE's off-axis angle relative to the BNB's on-axis flux.  

\subsection*{Neutrino flux simulation}

The simulation of the neutrino flux at MicroBooNE accounts for the production of hadrons from the initial interaction of the proton beam on the target and the propagation of these hadrons through a detailed beamline geometry description, achieved using the $\textsc{Geant4}$ toolkit~\cite{g4}.
Hadron production cross-sections are constrained by dedicated external measurements where available, tailored to the specific beam parameters, including target differences and initial proton beam energy.
The BNB simulation, identical to that used at MiniBooNE~\cite{MiniBooNE:2008hfu}, employs \textsc{Geant} \textnormal{v}4.10.4 with a custom physics list and constrains $\pi^{\pm}$ yields with data from the HARP experiment~\cite{harp}, along with an updated $K^{+}$ production constraint from SciBooNE~\cite{SciBooNE:2011sjq, Mariani:2011zd}. 
The NuMI simulation has been updated to \textsc{Geant} \textnormal{v}4.10.4 with the \textsc{FTFP-BERT} physics list~\cite{Bertini:1969rqi}.
The constraints on $\pi^{\pm}$ and $K^{\pm}$ yields from the NA49 experiment at CERN~\cite{na49k,na49pi} are implemented using the $\textsc{PPFX}$ toolkit~\cite{ppfx}, which has been updated to use the new $\textsc{Geant}$ version. 
Uncertainties are estimated for each process and various components of the beamline geometry, resulting in a combined systematic uncertainty of approximately 13\% for the BNB and 26\% for the NuMI beam on the integrated flux.
These uncertainties on the fluxes are different from the uncerztainties quoted in Tab.~\ref{tab:uncertainties} which are on the overall event rates.
These uncertainties are dominated by hadron production rates, and are larger for the NuMI beam because of the lack of constraints at large off-axis angles from dedicated hadron production experiments.
Since the nature of neutrino production in the two beams is different, the uncertainties are considered uncorrelated across the respective beams. 

\subsection*{Degeneracy between \nue appearance and disappearance}

The interplay between oscillation of intrinsic $\nu_\mu$ and $\nu_e$ components in the BNB and the NuMI beam is illustrated in Extended Data Fig.~\ref{extfig:Degeneracy}.
For various combinations of the expanded $4\nu$ PMNS mixing angles, and assuming $\Delta m^2_{41}=\unit[1.4]{eV^2}$, the ratio of predicted $\nu_e$ signal events with $0<E_\nu<\unit[2.5]{GeV}$ in MicroBooNE relative to the $3\nu$ prediction is shown for the BNB on the $x$ axis and for the NuMI beam on the $y$ axis.
By tracing vertically along $x = 1$, we observe that a MicroBooNE BNB \nue\ measurement could be consistent with the $3\nu$ hypothesis of $\theta_{ee}=\theta_{\mu e}=0$ as well as with non-zero mixing angles in the alternate $4\nu$ case.
The addition of a NuMI \nue\ measurement enables a much clearer interpretation of the allowed oscillation behaviour, while also strengthening the constraining power of the analysis.
Specifically, perfect agreement between data and the $3\nu$ prediction for both BNB and NuMI would favour the null oscillation case, while a large deficit in the high $\nu_e$-content NuMI beam would clearly indicate competing appearance and disappearance effects in the BNB $\nu_e$ sample. In addition, one can see that the range of allowed $4\nu$ predictions in NuMI and BNB when taken together are quite restricted, allowing us to set tighter limits on this oscillation model based on the observed CC $\nu_{e}$ event rates after constraints in Table~\ref{tab:uncertainties} ($x\sim0.9, y\sim1$).

Extended Data Fig.~\ref{extfig:Sensitivities} shows the impact of the degeneracy-breaking on the sensitivity to the $4\nu$ parameter space. The dashed lines show the exclusion regions of MicroBooNE's previous BNB-only analysis~\cite{WCOscPRL}, compared to the solid red lines which show the exclusions obtained by this analysis when including the NuMI beam data.

\subsection*{Statistical methods for oscillation analysis}

The combined statistical and systematic uncertainties on the 14 event samples are described by the covariance matrix
\begin{linenomath*}
\begin{equation}
    \Sigma = \left(
    \begin{array}{ccc}
    C_{1,1} & \cdots & C_{1,14} \\
    \vdots & \ddots & \vdots \\
    C_{14,1} & \cdots & C_{14,14}
    \end{array}
    \right),
\end{equation}
\end{linenomath*}
where the $C_{i,j}$ are the bin-by-bin covariance matrices between the $i$th and $j$th event samples. 
These covariance matrices are the sums of the covariance matrices arising from statistical uncertainties and from each source of systematic uncertainty,
\begin{linenomath*}
\begin{equation}
    C_{i,j}=C_{i,j}^\textrm{stat.}+\sum_{k}C_{i,j}^{\textrm{syst.}_k},
\end{equation}
\end{linenomath*}
where the the sum runs over the $k$ sources of systematic uncertainty.
The covariance matrix for the statistical uncertainty follows the Pearson format.

Figure~\ref{fig:data} and Tab.~\ref{tab:uncertainties} demonstrate the power of the 14 event samples to effectively constrain the systematic uncertainties due to the correlations present in the covariance matrix. 
To produce the constrained predictions shown in Fig.~\ref{fig:data} and the constrained systematic uncertainties in Tab.~\ref{tab:uncertainties}, a conditional constraint formalism~\cite{ref:EatonMultivariateStatistics} is employed, which uses the 14 event samples simultaneously to constrain the systematic uncertainties and to provide updated predictions for each event sample.
To understand how the constraint is applied to the $i$th event sample, the full unconstrained covariance matrix can be written as
\begin{linenomath*}
\begin{equation}\label{eq:covariance}
    \Sigma = \left(
    \begin{array}{cc}
    C_{i,i} & C_{i,x} \\
    C_{x,i} & C_{x,x}
    \end{array}
    \right),
\end{equation}
\end{linenomath*}
where elements with a subscript $x$ represent the remaining 13 blocks of the full matrix. An updated, constrained covariance matrix for the $i$th event sample is obtained as
\begin{linenomath*}
\begin{equation}\label{eq:conditional_cov}
    C_{i,i}^\textrm{constr.}=
    C_{i,i}-C_{i,x}\cdot\left(C_{x,x}\right)^{-1}\cdot C_{x,i}.
\end{equation}
\end{linenomath*}
Given the unconstrained binned prediction for the $i$th event sample, $\mu_i$, and the remaining binned prediction and data samples, $\mu_x$ and $n_x$, a constrained binned prediction can also be formed as
\begin{linenomath*}
\begin{equation}\label{eq:conditional_mean}
\mu_i^\textrm{constr.}=\mu_i+C_{i,x}\cdot\left(C_{x,x}\right)^{-1}\cdot\left(n_x-\mu_x\right).
\end{equation}
\end{linenomath*}
To perform the oscillation fit, a $\chi^2$ test statistic,
\begin{linenomath*}
\begin{equation}
    \chi^2=\left(N-M\right)^T\cdot\Sigma^{-1}\cdot\left(N-M\right),
\end{equation}
\end{linenomath*}
is formed using the 14 binned event samples from the data, $N=(n_1,\ldots,n_{14})$, where $n_i$ is the $i$th event sample from the data, and the corresponding unconstrained binned predictions for the 14 event samples, $M=(\mu_1\ldots\mu_{14})$.
The oscillation parameters used to produce the prediction are varied until $\chi^2$ is minimised.
Since the bin-to-bin correlations between the 14 event samples are contained in the matrix $\Sigma$, minimising this $\chi^2$ intrinsically incorporates the constraint procedure into the measurement of the oscillation parameters.
As $\chi^2$ is minimised the absolute systematic uncertainty varies as the number of oscillated neutrino interactions within the fiducial volume changes, while the fractional systematic uncertainty remains constant.
In contrast, the absolute systematic uncertainty related to non-neutrino backgrounds and out-of-fiducial-volume neutrino interactions remains unchanged. 
Consequently, the total covariance matrix is updated in accordance with the oscillation parameters of interest.

Exclusion limits on the oscillation parameters are calculated using the frequentist-motivated $\mathrm{CL_{s}}$ method~\cite{Read:2002hq}, which is commonly employed for determining exclusion limits in high-energy physics.
The $\mathrm{CL_{s}}$ test statistic is defined as $\mathrm{CL_{s}} = p_{4\nu}/p_{3\nu}$, where $p_{4\nu}$ and $p_{3\nu}$ are the one-sided $p$-values of $\Delta\chi^2_{\rm CL_s, data}$ under the $4\nu$ and the null $3\nu$ hypotheses, respectively, where
\[
\Delta \chi^2_{\rm CL_s}=\chi^2_{4\nu} - \chi^2_{3\nu}
\]
at a given point in the $4\nu$ parameter space.
These $p$-values are one-sided because the test statistic measures the deviation from the null in the specific direction of the alternate hypothesis.
The $p$-values are determined using a frequentist approach by generating pseudo-experiments with the full covariance matrix, assuming the respective hypothesis is true.
The region in which $\mathrm{CL}_{s} \leq 1 - \alpha$ is excluded at the confidence level $\alpha$.
By generating pseudo-experiments under the null $3\nu$ hypothesis, expected exclusion limits are calculated across the 2D parameter spaces of $\Delta m^2_{41}$ and $\sin^{2}(2\theta_{\mu e})$ or $\sin^{2}(2\theta_{ee})$.

To quantify the expected sensitivity of the analysis, the median $\mathrm{sin}^{2}(2\theta_{\mu e(ee)})$ value from all expected exclusion limits is determined for each $\Delta m^2_{41}$ value.
These median sensitivities are shown in Extended Data Fig.~\ref{extfig:Sensitivities}, and are compared to the exclusions set using the data.
To show the expected level of fluctuations of the measured limit from the median sensitivity, $1\sigma$ and $2\sigma$ bands are also shown in Extended Data Fig.~\ref{extfig:Sensitivities}, which encompass the central 68.3\% and 95.5\% of exclusions from the pseudo-experiments.
In $\sin^2(2\theta_{\mu e})$ space, our exclusion is 
stronger than our median expected sensitivity. 
Two factors contribute to this stronger exclusion. First, a deficit in the BNB CC \nue\ sample more strongly disfavors \nue\ appearance. Second, the excess in the NuMI CC \numu\ sample leads to a reduction in the constrained fractional uncertainty on the NuMI \nue\ prediction through the joint fit procedure, which in turn further strengthens the exclusion limit.
In $\sin^2(2\theta_{ee})$ space, the deficit in the BNB CC $\nu_e$ FC sample plays the opposite role, slightly favouring $\nu_e$ disappearance and making the exclusion contour weaker than the median sensitivity.

\subsection*{Impact of the NuMI CC \numu\ Sideband}
As shown in Tab. \ref{tab:uncertainties}, a combined fit of the 14 reconstructed samples constrains the signal CC \nue\ prediction and its uncertainties due to the correlations between the sideband and the signal channels. For the NuMI CC \nue\ signal sample in particular, a crucial driver of the constraint is the corresponding CC \numu sideband, shown in Extended Data Fig.~\ref{extfig:numi_numu}. Before the fit, the normalization difference between data and the prediction is $24.5\%$ with an overall uncertainty of $21.1\%$ on the prediction. 

In order to evaluate the impact of this difference on the combined fit, we can extend the covariance matrix used in the analysis (364$\times$364, corresponding to the energy spectra of the 14 channels) by adding an additional bin representing the overall NuMI CC \numu\ normalization (combining FC and PC) and computing the respective covariances with the other 364 analysis bins, resulting in a 365$\times$365 matrix. One can then obtain a post-fit mean and error on this normalization parameter by constraining the 364$\times$364 block to the data using Eqns. \eqref{eq:conditional_cov} and \eqref{eq:conditional_mean}. This gives us an estimate of how much this parameter is effectively being pulled in the combined fit. The post-fit value of this parameter is $1.28 \pm 0.058$, indicating consistency with the corresponding observed value as well as a modest pull of $\sim 1.3\sigma$.


\clearpage
\onecolumn
\section*{The MicroBooNE Collaboration}

P.~Abratenko\textsuperscript{39},
D.~Andrade~Aldana\textsuperscript{14},
L.~Arellano\textsuperscript{22},
J.~Asaadi\textsuperscript{38},
A.~Ashkenazi\textsuperscript{37},
S.~Balasubramanian\textsuperscript{12},
B.~Baller\textsuperscript{12},
A.~Barnard\textsuperscript{29},
G.~Barr\textsuperscript{29},
D.~Barrow\textsuperscript{29}
% ...

\vspace{0.5cm}

\noindent\textsuperscript{1} Argonne National Laboratory (ANL), Lemont, IL, 60439, USA\\
\noindent\textsuperscript{2} Universit{\"a}t Bern, Bern CH-3012, Switzerland\\
\noindent\textsuperscript{3} Brookhaven National Laboratory (BNL), Upton, NY, 11973, USA\\
\noindent\textsuperscript{4} University of California, Santa Barbara, CA, 93106, USA\\
\noindent\textsuperscript{5} University of Cambridge, Cambridge CB3 0HE, United Kingdom\\
\noindent\textsuperscript{6} Centro de Investigaciones Energ\'{e}ticas, Medioambientales y Tecnol\'{o}gicas (CIEMAT), Madrid E-28040, Spain\\
\noindent\textsuperscript{7} University of Chicago, Chicago, IL, 60637, USA\\
\noindent\textsuperscript{8} University of Cincinnati, Cincinnati, OH, 45221, USA\\
\noindent\textsuperscript{9} Colorado State University, Fort Collins, CO, 80523, USA\\
\noindent\textsuperscript{10} Columbia University, New York, NY, 10027, USA\\
\noindent\textsuperscript{11} University of Edinburgh, Edinburgh EH9 3FD, United Kingdom\\
\noindent\textsuperscript{12} Fermi National Accelerator Laboratory (FNAL), Batavia, IL 60510, USA\\
\noindent\textsuperscript{13} Universidad de Granada, E-18071, Granada, Spain\\
\noindent\textsuperscript{14} Illinois Institute of Technology (IIT), Chicago, IL 60616, USA\\
\noindent\textsuperscript{15} Imperial College London, London SW7 2AZ, United Kingdom\\
\noindent\textsuperscript{16} Indiana University, Bloomington, IN 47405, USA\\
\noindent\textsuperscript{17} The University of Kansas, Lawrence, KS, 66045, USA\\
\noindent\textsuperscript{18} Kansas State University (KSU), Manhattan, KS, 66506, USA\\
\noindent\textsuperscript{19} Lancaster University, Lancaster LA1 4YW, United Kingdom\\
\noindent\textsuperscript{20} Los Alamos National Laboratory (LANL), Los Alamos, NM, 87545, USA\\
\noindent\textsuperscript{21} Louisiana State University, Baton Rouge, LA, 70803, USA\\
\noindent\textsuperscript{22} The University of Manchester, Manchester M13 9PL, United Kingdom\\
\noindent\textsuperscript{23} Massachusetts Institute of Technology (MIT), Cambridge, MA, 02139, USA\\
\noindent\textsuperscript{24} University of Michigan, Ann Arbor, MI, 48109, USA\\
\noindent\textsuperscript{25} Michigan State University, East Lansing, MI 48824, USA\\
\noindent\textsuperscript{26} University of Minnesota, Minneapolis, MN, 55455, USA\\
\noindent\textsuperscript{27} Nankai University, Nankai District, Tianjin 300071, China\\
\noindent\textsuperscript{28} New Mexico State University (NMSU), Las Cruces, NM, 88003, USA\\
\noindent\textsuperscript{29} University of Oxford, Oxford OX1 3RH, United Kingdom\\
\noindent\textsuperscript{30} University of Pittsburgh, Pittsburgh, PA, 15260, USA\\
\noindent\textsuperscript{31} Queen Mary University of London, London E1 4NS, United Kingdom\\
\noindent\textsuperscript{32} Rutgers University, Piscataway, NJ, 08854, USA\\
\noindent\textsuperscript{33} SLAC National Accelerator Laboratory, Menlo Park, CA, 94025, USA\\
\noindent\textsuperscript{34} South Dakota School of Mines and Technology (SDSMT), Rapid City, SD, 57701, USA\\
\noindent\textsuperscript{35} University of Southern Maine, Portland, ME, 04104, USA\\
\noindent\textsuperscript{36} Syracuse University, Syracuse, NY, 13244, USA\\
\noindent\textsuperscript{37} Tel Aviv University, Tel Aviv, Israel, 69978\\
\noindent\textsuperscript{38} University of Texas, Arlington, TX, 76019, USA\\
\noindent\textsuperscript{39} Tufts University, Medford, MA, 02155, USA\\
\noindent\textsuperscript{40} Center for Neutrino Physics, Virginia Tech, Blacksburg, VA, 24061, USA\\
\noindent\textsuperscript{41} University of Warwick, Coventry CV4 7AL, United Kingdom\\

\clearpage
\newpage
\section*{Extended Data}
\setcounter{figure}{0}

\makeatletter

\renewcommand{\theHfigure}{ED\arabic{figure}}
\makeatother

\begin{figure}[ht]
    \centering
    \includegraphics[width=0.5\textwidth]{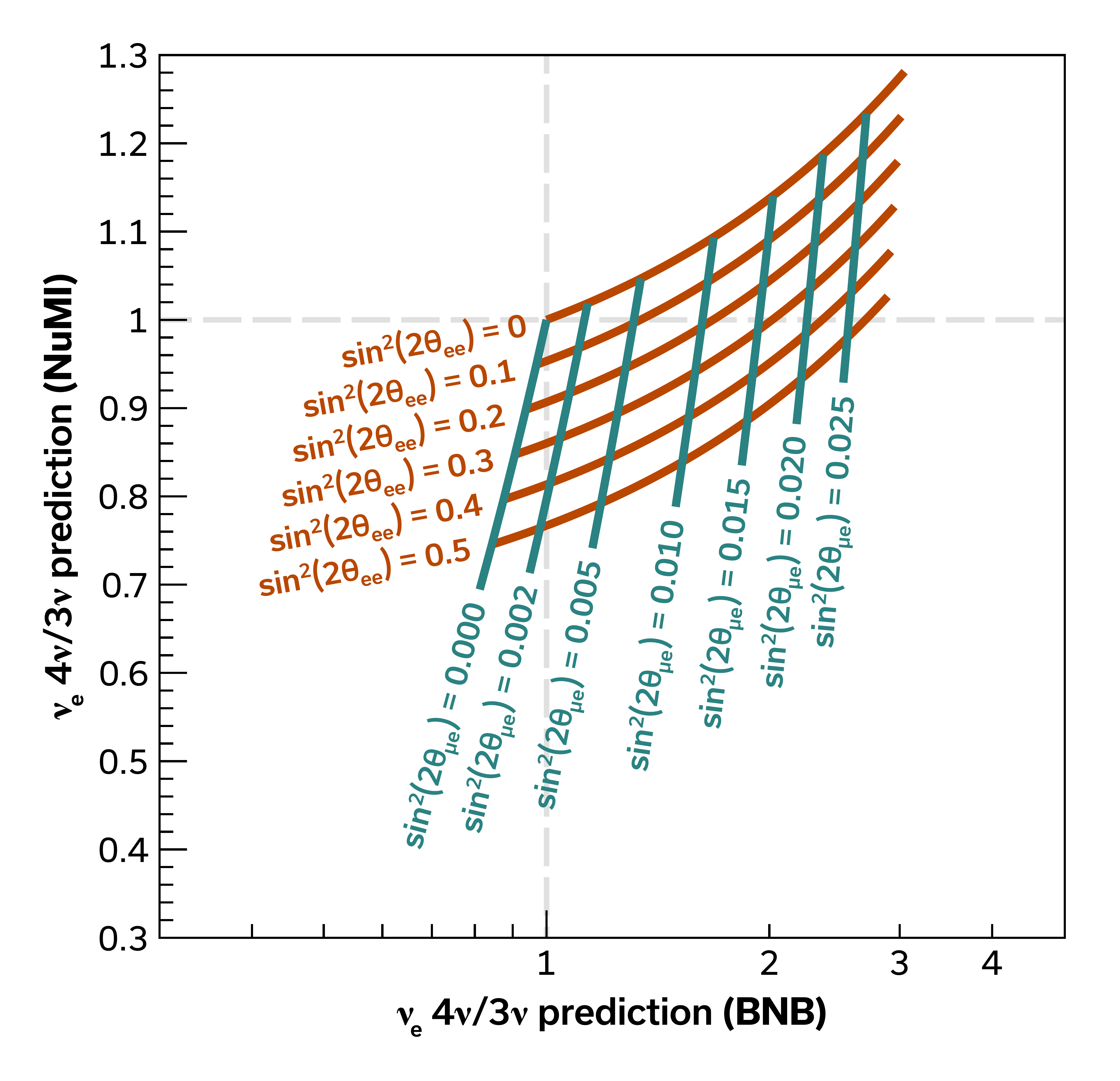}
    \caption{{\bf $\nu_e$ event rates in MicroBooNE as a function of oscillation parameters.} The $x$-axis (in log scale) shows the expected CC $\nu_e$ event rate in the MicroBooNE detector in the $4\nu$ scenario, with respect to the $3\nu$ scenario, from the BNB. The $y$-axis (in linear scale) shows the same ratio from the NuMI beam. The lines indicate how these ratios depend on the oscillation parameters of the expanded $4\nu$ PMNS matrix for $\Delta m^2_{41}=\unit[1.4]{eV^2}$.}
    \label{extfig:Degeneracy}
\end{figure}

\begin{figure}[ht]
	\centering
	\includegraphics[trim={0 3cm 0 55cm},clip,width=0.5\textwidth]{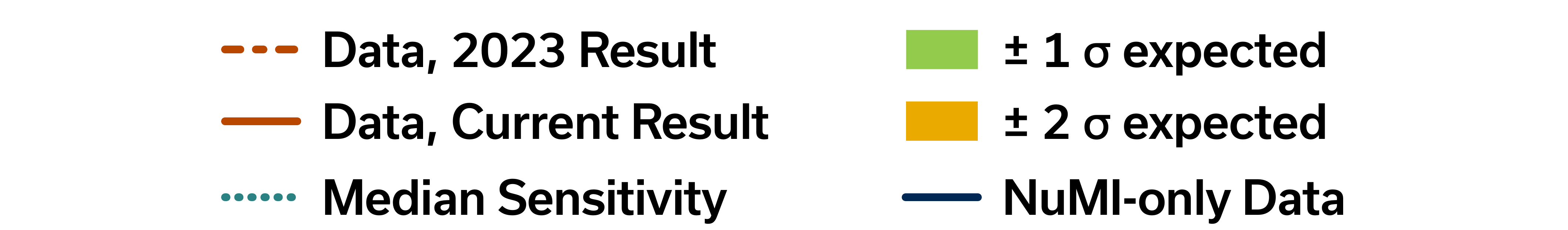}
	\begin{overpic}[trim={0 14cm 0 2cm},clip,width=0.5\textwidth]{SuppFig2a.pdf}
		\put(180,670){\textbf{a}}
	\end{overpic} 
	\begin{overpic}[trim={0 14cm 0 2cm},clip,width=0.5\textwidth]{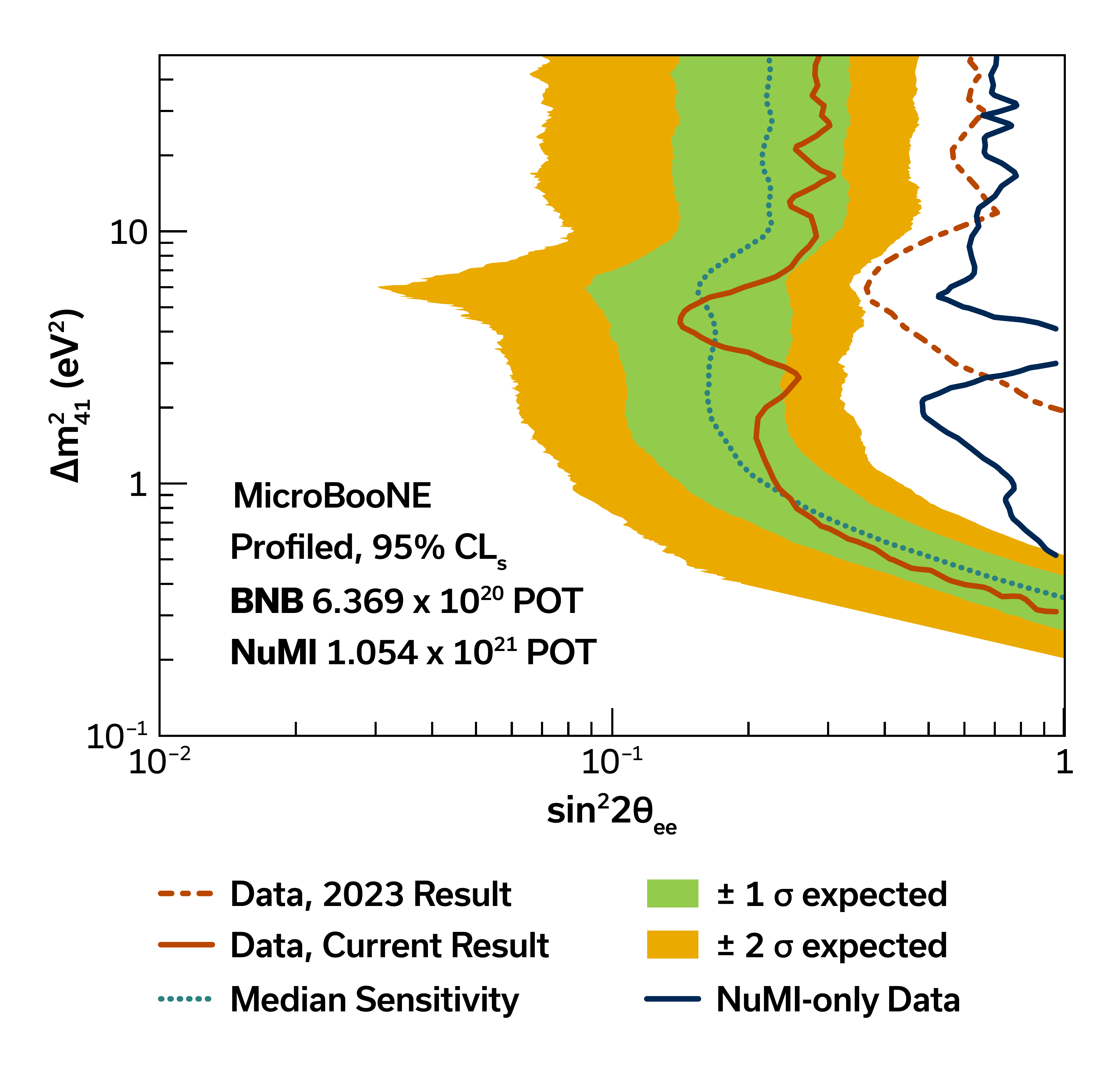}
		\put(180,670){\textbf{b}}
	\end{overpic}
	\caption{{\bf Comparison of the measured exclusions to the expected sensitivities.}
		The red solid lines show the measured exclusion lines at the 95\% CL$_\text{s}$ level in the plane of $\Delta m^2_{41}$ and ({\bf a}) $\sin^2(2\theta_{\mu e})$ or ({\bf b}) $\sin^{2}(2\theta_{ee})$.
		The blue dashed lines show the median expected sensitivities.
		The green and yellow bands show the $1\sigma$ and $2\sigma$ expected fluctuations around the median sensitivities at each $\Delta m^2_{41}$ value.
		The red dashed lines show the previous MicroBooNE BNB-only result~\cite{WCOscPRL} and the blue solid lines shows the measured NuMI-only exclusions at 95\% CL$_\text{s}$.} 
	\label{extfig:Sensitivities}
\end{figure}

\begin{figure}[ht]
    \centering
    \begin{overpic}[width=0.5\textwidth]{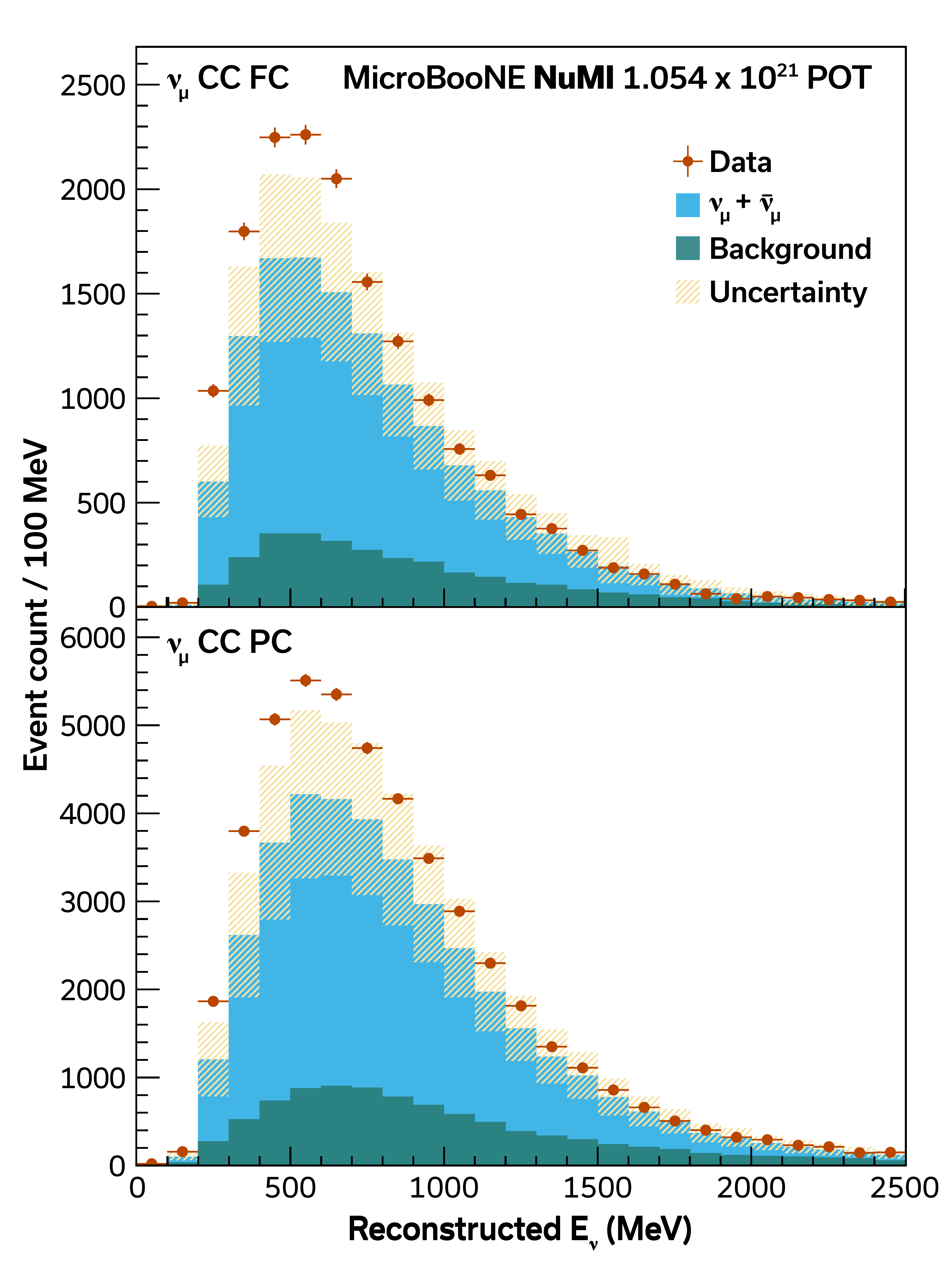}
         \put(135,875){\bf a}
         \put(135,435){\bf b}
    \end{overpic}
    \caption{{\bf Observed NuMI CC \numu candidate events.} 
    Reconstructed energy spectra of events selected as ({\bf a}) fully contained CC $\nu_\mu$ candidates in the NuMI beam, and ({\bf b}) partially contained CC $\nu_\mu$ candidates in the NuMI beam.
    The data points are shown with statistical error bars.
    The constrained predictions for each sample are shown for the $3\nu$ hypothesis as the solid histograms, with the blue showing the true CC $\nu_\mu$ events and the green showing the background events.
    The background category contains NC neutrino interactions, cosmic rays, CC $\nu_e$ interactions, and interactions occurring outside the fiducial volume of the detector.
    The yellow band shows the total systematic uncertainty on the prediction.}
    \label{extfig:numi_numu}
\end{figure}

\end{document}